\documentclass[twoside]{article}%
\usepackage{amssymb}
\usepackage{amsfonts}
\usepackage{amsmath}
\usepackage{graphicx}%
\providecommand{\U}[1]{\protect\rule{.1in}{.1in}}
\newtheorem{theorem}{Theorem}

\newtheorem{lemma}[theorem]{Lemma}

\newtheorem{remark}[theorem]{Remark}

\newenvironment{proof}[1][Proof]{\noindent\textbf{#1.} }{\ \rule{0.5em}{0.5em}}
\ifx\pdfoutput\relax\let\pdfoutput=\undefined\fi
\newcount\msipdfoutput
\ifx\pdfoutput\undefined\else
\ifcase\pdfoutput\else
\ifx\paperwidth\undefined\else
\ifdim\paperheight=0pt\relax\else\pdfpageheight\paperheight\fi
\ifdim\paperwidth=0pt\relax\else\pdfpagewidth\paperwidth\fi
\fi\fi\fi
\begin{document}

\title{\textbf{Analytical Solutions to the Navier-Stokes-Poisson Equations with
Density-dependent Viscosity and with Pressure}}
\author{\textsc{Ling Hei Yeung\thanks{E-mail address: lightisgood2005@yahoo.com.hk}}
\and M\textsc{anwai Yuen\thanks{E-mail address: nevetsyuen@hotmail.com }}\\\textit{Department of Applied Mathematics, The Hong Kong Polytechnic
University,}\\\textit{Hung Hom, Kowloon, Hong Kong}}
\date{Revised 05-May-2010}
\maketitle

\begin{abstract}
We study some particular solutions to the Navier-Stokes-Poisson equations with
density-dependent viscosity and with pressure, in radial symmetry. With
extension of the previous known blowup solutions for the Euler-Poisson
equations / pressureless Navier-Stokes-Poisson with density-dependent
viscosity, we constructed the corresponding analytical blowup solutions for
the Navier-Stokes-Poisson Equations with density-dependent viscosity and with pressure.


\end{abstract}

\section{Introduction}

The evolution of a self-gravitating fluid can be formulated by the isentropic
Euler-Poisson equations of the following form:
\begin{equation}
\left\{
\begin{array}
[c]{rl}%
{\normalsize \rho}_{t}{\normalsize +\nabla\cdot(\rho\vec{u})} &
{\normalsize =}{\normalsize 0,}\\
{\normalsize \rho\lbrack\vec{u}}_{t}+\left(  {\normalsize \vec{u}\cdot\nabla
}\right)  {\normalsize \vec{u})]+\nabla P} & {\normalsize =-}%
{\normalsize \delta\rho\nabla\Phi+vis(\rho,\vec{u}),}\\
{\normalsize \Delta\Phi(t,x)} & {\normalsize =\alpha(N)}{\normalsize \rho
-\Lambda,}%
\end{array}
\right.  \label{Euler-Poisson}%
\end{equation}
where $\alpha(N)$ is a constant related to the unit ball in $R^{N}$:
$\alpha(1)=2$; $\alpha(2)=2\pi$ and for $N\geq3,$%
\begin{equation}
\alpha(N)=N(N-2)Vol(N)=N(N-2)\frac{\pi^{N/2}}{\Gamma(N/2+1)},
\end{equation}
where $Vol(N)$ is the volume of the unit ball in $R^{N}$ and $\Gamma$ is a
Gamma function. And as usual, $\rho=\rho(t,\vec{x})$ and $\vec{u}=\vec
{u}(t,\vec{x})\in\mathbf{R}^{N}$ are the density and the velocity
respectively. $P=P(\rho)$\ is the pressure. And ${\normalsize \Lambda}$ is the
background constant.

When $\delta=1$, the system can model fluids that are self-gravitating , such
as gaseous stars. In addition, the evolution of the simple cosmology can be
modelled by the dust distribution without pressure term. This describes the
stellar systems of collisionless and gravitational $n$-body systems \cite{FT}.
And the pressureless Euler-Poisson equations can be derived from the
Vlasov-Poisson-Boltzmann model with the zero mean free path \cite{G1}. For
$N=3$, the equations (\ref{Euler-Poisson}) are the classical
(non-relativistic) descriptions of a galaxy in astrophysics. See \cite{BT} and
\cite{C}, for details about the systems.\newline When $\delta=-1$, the system
is the compressible Euler-Poisson equations with repulsive forces. The
equation (\ref{Euler-Poisson})$_{3}$ is the Poisson equation through which the
potential with repulsive forces is determined by the density distribution of
the electrons. In this case, the system can be viewed as a semiconductor
model. See \cite{Cse} and \cite{Lions} for detailed analysis of the
system.\newline When $\delta=0$, the potential forces are ignored. The system
is called Euler / Navier-Stokes equations. See \cite{CW}, \cite{Ni} and
\cite{T} for detailed analysis of the system

In the above system, the self-gravitational potential field $\Phi=\Phi
(t,x)$\ is determined by the density $\rho$ through the Poisson equation.

And $vis(\rho,\vec{u})$ is the viscosity function:%
\begin{equation}
vis(\rho,\vec{u}):=\bigtriangledown(\mu(\rho)\bigtriangledown\cdot\vec{u}).
\label{pp1}%
\end{equation}
In this article, we seek the radial solutions
\begin{equation}
\rho(t,\vec{x})=\rho(t,r)\text{ and }\vec{u}=\frac{\vec{x}}{r}V(t,r)=:\frac
{\vec{x}}{r}V\text{,}%
\end{equation}
with $r=\left(  \sum_{i=1}^{N}x_{i}^{2}\right)  ^{1/2}$. By the standard
computation, the Euler-Poisson equations in radial symmetry can be written in
the following form:%
\begin{equation}
\left\{
\begin{array}
[c]{c}%
\rho_{t}+V\rho_{r}+\rho V_{r}+\dfrac{N-1}{r}\rho V=0,\\
\rho\left(  V_{t}+VV_{r}\right)  +P_{r}(\rho)+{\normalsize \delta}\rho\Phi
_{r}\left(  \rho\right)  =\mu_{r}\left(  \rho\right)  \left(  \frac{N-1}%
{r}V+V_{r}\right)  +\mu(\rho)\left[  -\left(  \frac{N-1}{r^{2}}\right)
V+\frac{N-1}{r}V_{r}+V_{rr}\right]  .
\end{array}
\right.  \label{gooo}%
\end{equation}
Under a common assumption for, the viscosity function can be defined:
\begin{equation}
\mu(\rho):=\kappa\rho^{\theta}, \label{fdfd}%
\end{equation}
where $\kappa$ and $\theta\geq0$ are the constants. For the study of the above
system, the readers may refer \cite{LL}, \cite{Ni}, \cite{YZ} and \cite{Y1}.
In particular, when $\theta=0$, that returns the expression for the only
$V$-dependent viscosity function:
\begin{equation}
\left\{
\begin{array}
[c]{c}%
\rho_{t}+V\rho_{r}+\rho V_{r}+\dfrac{N-1}{r}\rho V=0,\\
\rho\left(  V_{t}+VV_{r}\right)  +P_{r}(\rho)+{\normalsize \delta}\rho\Phi
_{r}\left(  \rho\right)  =\kappa\left[  V_{rr}+\frac{N-1}{r}V_{r}-\left(
\frac{N-1}{r^{2}}\right)  V\right]  ,
\end{array}
\right.
\end{equation}
which are the usual form of the Navier-Stokes-Poisson equations. The equations
(\ref{Euler-Poisson})$_{1}$ and (\ref{Euler-Poisson})$_{2}$ $(vis(\rho
,u)\neq0)$ are the compressible Navier-Stokes equations with forcing term. The
equation (\ref{Euler-Poisson})$_{3}$ is the Poisson equation through which the
gravitational potential is determined by the density distribution of the
density itself. Thus, the system (\ref{Euler-Poisson}) is called the
Navier-Stokes-Poisson equations.

Here, if the $vis(\rho,u)=0$, the system is called the Euler-Poisson
equations.\ In this case, the equations can be viewed as a prefect gas model.
For $N=3$, (\ref{Euler-Poisson}) is a classical (nonrelativistic) description
of a galaxy, in astrophysics. See \cite{C}, \cite{M1} for details about the system.

$P=P(\rho)$\ is the pressure. The $\gamma$-law can be applied on the pressure
$P(\rho)$, i.e.%
\begin{equation}
{\normalsize P}\left(  \rho\right)  {\normalsize =K\rho}^{\gamma}%
:=\frac{{\normalsize \rho}^{\gamma}}{\gamma}, \label{gamma}%
\end{equation}
which is a common hypothesis. The constant $\gamma=c_{P}/c_{v}\geq1$, where
$c_{P}$, $c_{v}$\ are the specific heats per unit mass under constant pressure
and constant volume respectively, is the ratio of the specific heats, that is,
the adiabatic exponent in (\ref{gamma}). With $K=0$, we call that the system
is pressureless.

For the local existence of the Euler-Poisson equations, the interested reader
may refer the results of Makino \cite{M}, Gambin \cite{G} and Bezard \cite{B}.

In the following, we always seek solutions in radial symmetry. Thus, the
Poisson equation (\ref{Euler-Poisson})$_{3}$ is transformed to%
\begin{equation}
{\normalsize r^{N-1}\Phi}_{rr}\left(  {\normalsize t,x}\right)  +\left(
N-1\right)  r^{N-2}\Phi_{r}{\normalsize =}\alpha\left(  N\right)
{\normalsize \rho r^{N-1},}%
\end{equation}%
\begin{equation}
\Phi_{r}=\frac{\alpha\left(  N\right)  }{r^{N-1}}\int_{0}^{r}\left(
\rho(t,s)-\Lambda\right)  s^{N-1}ds.
\end{equation}
In this paper, we concern about blowup solutions for the $N$-dimensional
pressureless Navier-Stokes-Poisson equations with the density-dependent
viscosity. And our aim is to construct a family of blowup solutions.

Historically in astrophysics, Goldreich and Weber \cite{GW} constructed the
analytical blowup (collapsing) solutions of the $3$-dimensional Euler-Poisson
equations for $\gamma=4/3$ and for the non-rotating gas spheres. After that,
Makino \cite{M1} obtained the rigorously mathematical proof of the existence
of such kind of blowup solutions. And in \cite{DXY}, Deng, Xiang and Yang
extended the above blowup solutions in $R^{N}$ ($N\geq3$). Recently, Yuen
obtained the blowup solutions in $R^{2}$ with $\gamma=1$ in \cite{Y1}. The
family of the analytical solutions are rewritten as:\newline for $N\geq3$ and
$\gamma=(2N-2)/N$, in \cite{DXY}:
\begin{equation}
\left\{
\begin{array}
[c]{c}%
\rho(t,r)=\left\{
\begin{array}
[c]{c}%
\dfrac{1}{a(t)^{N}}y(\frac{r}{a(t)})^{N/(N-2)},\text{ for }r<a(t)Z_{\mu};\\
0,\text{ for }a(t)Z_{\mu}\leq r.
\end{array}
\right.  \text{, }V{\normalsize (t,r)=}\dfrac{\dot{a}(t)}{a(t)}%
{\normalsize r,}\\
\ddot{a}(t){\normalsize =}\dfrac{-\lambda}{a(t)^{N-1}},\text{ }%
{\normalsize a(0)=a}_{0}>0{\normalsize ,}\text{ }\dot{a}(0){\normalsize =a}%
_{1},\\
\ddot{y}(z){\normalsize +}\dfrac{N-1}{z}\dot{y}(z){\normalsize +}\dfrac
{\alpha(N)}{(2N-2)K}{\normalsize y(z)}^{N/(N-2)}{\normalsize =\mu,}\text{
}y(0)=\alpha>0,\text{ }\dot{y}(0)=0,
\end{array}
\right.  \label{solution2}%
\end{equation}
where $\mu=[N(N-2)\lambda]/(2N-2)K$ and the finite $Z_{\mu}$ is the first zero
of $y(z)$;\newline for $N=2$ and $\gamma=1$, in \cite{Y1}%
\begin{equation}
\left\{
\begin{array}
[c]{c}%
\rho(t,r)=\dfrac{1}{a(t)^{2}}e^{y(r/a(t))}\text{, }V{\normalsize (t,r)=}%
\dfrac{\dot{a}(t)}{a(t)}{\normalsize r,}\\
\ddot{a}(t){\normalsize =}\dfrac{-\lambda}{a(t)},\text{ }{\normalsize a(0)=a}%
_{0}>0{\normalsize ,}\text{ }\dot{a}(0){\normalsize =a}_{1},\\
\ddot{y}(z){\normalsize +}\dfrac{1}{z}\dot{y}(z){\normalsize +\dfrac{2\pi}%
{K}e}^{y(z)}{\normalsize =\mu,}\text{ }y(0)=\alpha,\text{ }\dot{y}(0)=0,
\end{array}
\right.  \label{solution 3}%
\end{equation}
where $K>0$, $\mu=2\lambda/K$ with a sufficiently small $\lambda$ and $\alpha$
are constants.\newline For the construction of analytical solutions to the
Euler, Navier-Stokes and pressureless Navier-Stokes-Poisson equations, readers
may refer to the recent results in \cite{Y}, \cite{YY}, \cite{Y3} and
\cite{Y4}.

In this article, it is natural to extend the more general results to cover the
full system (\ref{gooo}) ---- Navier-Stokes-Poisson equations with
density-dependent viscosity and \textit{with pressure}. In short, we
successfully deduce the system (\ref{gooo}) into the much simpler ordinary
differential equations.

\begin{theorem}
\label{thm:1}For the Navier-Stokes-Poisson equations in $R^{N}$ $(N\geq2)$
with $\gamma=\frac{2N-2}{N},$ (\ref{gooo}) there exists a family of solutions,
those are:%
\begin{equation}
\left\{  \rho(t,r)=\left\{
\begin{array}
[c]{c}%
\frac{f(\frac{r}{a(t)})}{a(t)^{N}},\text{ for }r<a(t)Z_{\mu},\\
0,\text{ for }a(t)Z_{\mu}\leq r,
\end{array}
\right.  ,\text{ }{\normalsize V}(t,r)=\frac{\dot{a}(t)}{a(t)}r\right.
,\label{yy111}%
\end{equation}
where $f(z)$ and $a(t)$ are the following functions and the finite $Z_{\mu}$
is the first zero of $f(z)$:\newline(1) with $\Lambda=0$, and\newline(1a)
$\theta=\frac{2N-3}{N}$,
\begin{equation}
\left\{
\begin{array}
[c]{c}%
a(t)=mt+n,\\
\gamma Kf(z)^{\gamma-2}\dot{f}(z)-m\kappa\theta Nf(z)^{\theta-2}\dot
{f}(z)+\frac{\delta\alpha(N)}{z^{N-1}}\int_{0}^{z}f(s)s^{N-1}ds,\text{
}f(0)=\alpha>0,
\end{array}
\right.  \label{yy}%
\end{equation}
where $m,$ $n>0$ and $\alpha$ are constant;\newline(1b) $\theta=\frac
{3N-4}{2N}$,
\begin{equation}
\left\{
\begin{array}
[c]{c}%
a(t)=(mt+n)^{2/N},\\
\gamma Kf(z)^{\gamma-2}\dot{f}(z)-2m\kappa\theta f(z)^{\theta-2}\dot
{f}(z)+\frac{\delta\alpha(N)}{z^{N-1}}\int_{0}^{z}f(s)s^{N-1}ds=\frac
{2(N-2)m^{2}}{N^{2}}z,\text{ }f(0)=\alpha>0.
\end{array}
\right.  ;\label{yy2}%
\end{equation}
In particular, for $m<0$, the solutions (\ref{yy}) and (\ref{yy2}) blow up at
the finite time $T=-m/n$.\newline(2) with $\delta\Lambda>0$ and $\theta
=\frac{2N-3}{N},$ we have%
\begin{equation}
\left\{
\begin{array}
[c]{c}%
a(t)=e^{\sqrt{\frac{\delta\alpha(N)\Lambda}{N}}t},\\
\left(  K\gamma-N\delta\alpha(N)\Lambda\right)  f(z)^{\theta-2}\dot
{f}(z)+\frac{\delta\alpha(N)}{z^{N-1}}\int_{0}^{z}f(s)s^{N-1}ds,\text{
}f(0)=\alpha,\text{ }f(0)=\alpha>0.
\end{array}
\right.
\end{equation}

\end{theorem}

\section{Separable Solutions}

Before we present the proof of Theorem \ref{thm:1}, Lemmas 3 and 12 of
\cite{Y2} are quoted here.

\begin{lemma}
[Lemma 3 of \cite{Y2}]\label{lem:generalsolutionformasseq copy(1)}For the
equation of conservation of mass in radial symmetry:
\begin{equation}
\rho_{t}+V\rho_{r}+\rho V_{r}+\frac{N-1}{r}\rho V=0,
\end{equation}
there exist solutions,%
\begin{equation}
\rho(t,r)=\frac{f(r/a(t))}{a(t)^{N}},\text{ }V{\normalsize (t,r)=}%
\frac{\overset{\cdot}{a}(t)}{a(t)}{\normalsize r,}%
\end{equation}
with the form $f\geq0\in C^{1}$ and $a(t)>0\in C^{1}.$
\end{lemma}

It is clear to check our solutions to satisfy the Navier-Stokes-Poisson equations:

\begin{proof}
As we use the functional structure of the above lemma, our solutions
(\ref{yy111}) fit well to the mass equation (\ref{gooo})$_{1}$.\newline(1a)
For the momentum equation (\ref{gooo})$_{2}$, we have:%
\begin{align}
&  \rho(V_{t}+V\cdot V_{r})+K\left(  \rho^{\gamma}\right)  _{r}-\kappa
{\normalsize (}\rho^{\theta})_{r}\left(  \frac{N-1}{r}V+V_{r}\right)
-\kappa\rho_{r}(V_{rr}+\frac{N-1}{r}V_{r}-\frac{N-1}{r^{2}}V)\\[0.1in]
&  +\rho\frac{\delta\alpha(N)}{r^{N-1}}\int_{0}^{r}\rho s^{N-1}ds\\[0.1in]
&  =\rho\frac{\ddot{a}(t)}{a(t)}r+K\left(  \left[  \frac{f(\frac{r}{a(t)}%
)}{a(t)^{N}}\right]  ^{\gamma}\right)  _{r}-\kappa\left(  \left[
\frac{f(\frac{r}{a(t)})}{a(t)^{N}}\right]  ^{\theta}\right)  _{r}\frac
{N\dot{a}(t)}{a(t)}+\rho\frac{\delta\alpha(N)}{r^{N-1}}\int_{0}^{r}%
\frac{f(\frac{r}{a(t)})}{a(t)^{N}}s^{N-1}ds\\[0.1in]
&  =\gamma K\left[  \frac{f(\frac{r}{a(t)})}{a(t)^{N}}\right]  _{{}}%
^{\gamma-1}\frac{\partial}{\partial r}\left[  \frac{f(\frac{r}{a(t)}%
)}{a(t)^{N}}\right]  -\kappa\theta\left[  \frac{f(\frac{r}{a(t)})}{a(t)^{N}%
}\right]  ^{\theta-1}\frac{\partial}{\partial r}\left[  \frac{f(\frac{r}%
{a(t)})}{a(t)^{N}}\right]  \frac{N\dot{a}(t)}{a(t)}\\[0.1in]
&  +\rho\frac{\delta\alpha(N)}{r^{N-1}}\int_{0}^{r}\frac{f(\frac{r}{a(t)}%
)}{a(t)^{N}}s^{N-1}ds,
\end{align}
with $a(t)=mt+n.$\newline Then, we have:%
\begin{align}
&  =\gamma K\left[  \frac{f(\frac{r}{a(t)})}{a(t)^{N}}\right]  _{{}}%
^{\gamma-1}\frac{\dot{f}(\frac{r}{a(t)})}{a(t)^{N+1}}-\kappa\theta\left[
\frac{f(\frac{r}{a(t)})}{a(t)^{N}}\right]  ^{\theta-1}\frac{\dot{f}(\frac
{r}{a(t)})}{a(t)^{N+2}}N\dot{a}(t)+\rho\frac{\delta\alpha(N)}{r^{N-1}}\int
_{0}^{r}\frac{f(\frac{r}{a(t)})}{a(t)^{N}}s^{N-1}ds\\[0.1in]
&  =\rho\left\{  \frac{\gamma Kf(\frac{r}{a(t)})^{\gamma-2}\dot{f}(\frac
{r}{a(t)})}{a(t)^{N(\gamma-2)+N+1}}-\frac{\kappa\theta Nf(\frac{r}%
{a(t)})^{\theta-2}\dot{f}(\frac{r}{a(t)})}{a(t)^{N\left(  \theta-2\right)
+N+2}}\dot{a}(t)+\frac{\delta\alpha(N)}{r^{N-1}}\int_{0}^{r}\frac{f(\frac
{r}{a(t)})}{a(t)^{N}}s^{N-1}ds\right\}  \\[0.1in]
&  =\frac{\rho}{a(t)^{N-1}}\left\{  \frac{\gamma Kf(\frac{r}{a(t)})^{\gamma
-2}\dot{f}(\frac{r}{a(t)})}{a(t)^{N(\gamma-2)+2}}-\frac{\kappa\theta
Nf(\frac{r}{a(t)})^{\theta-2}\dot{f}(\frac{r}{a(t)})}{a(t)^{N\left(
\theta-2\right)  +3}}\dot{a}(t)+\frac{\delta\alpha(N)}{\frac{r^{N-1}%
}{a(t)^{N-1}}}\int_{0}^{r/a(t)}\frac{f(s)}{a(t)^{N}}s^{N-1}ds\right\}  .
\end{align}
And we plug with the condition for $\gamma=\frac{2N-2}{N}$ and $\theta
=\frac{2N-3}{N}$ and with the new variable $z:=r/a(t)$:%
\begin{equation}
=\frac{\rho}{a(t)^{N-1}}\left\{  \gamma Kf(z)^{\gamma-2}\dot{f}(z)-m\kappa
\theta Nf(z)^{\theta-2}\dot{f}(z)+\frac{\delta\alpha(N)}{z^{N-1}}\int_{0}%
^{z}f(s)s^{N-1}ds\right\}  .
\end{equation}
In the theorem, we require the ordinary differential equation for $y(z)$ is
\begin{equation}
\left\{
\begin{array}
[c]{c}%
\gamma Kf(z)^{\gamma-2}\dot{f}(z)-m\kappa\theta Nf(z)^{\theta-2}\dot
{f}(z)+\frac{\delta\alpha(N)}{z^{N-1}}\int_{0}^{z}f(s)s^{N-1}ds=0,\\
y(0)=\alpha>0.
\end{array}
\right.
\end{equation}
Therefore, our solutions satisfy the momentum equation.

(1a) Similarly, for $\Lambda=0,$ $\gamma=\frac{2N-2}{N}$ and $\theta
=\frac{3N-4}{2N}$, we have:%
\begin{align}
&  \rho(V_{t}+V\cdot V_{r})+K\left(  \rho^{\gamma}\right)  _{r}-\kappa
{\normalsize (}\rho^{\theta})_{r}\left(  \frac{N-1}{r}V+V_{r}\right)
-\kappa\rho_{r}(V_{rr}+\frac{N-1}{r}V_{r}-\frac{N-1}{r^{2}}V)\\[0.1in]
&  +\rho\frac{\delta\alpha(N)}{r^{N-1}}\int_{0}^{r}\rho s^{N-1}ds\\[0.1in]
&  =\rho\frac{\ddot{a}(t)}{a(t)}r+K\left(  \left[  \frac{f(\frac{r}{a(t)}%
)}{a(t)^{N}}\right]  ^{\gamma}\right)  _{r}-\kappa\left(  \left[
\frac{f(\frac{r}{a(t)})}{a(t)^{N}}\right]  ^{\theta}\right)  _{r}\frac
{N\dot{a}(t)}{a(t)}+\rho\frac{\delta\alpha(N)}{r^{N-1}}\int_{0}^{r}%
\frac{f(\frac{r}{a(t)})}{a(t)^{N}}s^{N-1}ds\\[0.1in]
&  =\rho\frac{\frac{2}{N}(\frac{2}{N}-1)m^{2}}{(mt+n)^{2}}r+\gamma K\left[
\frac{f(\frac{r}{a(t)})}{a(t)^{N}}\right]  _{{}}^{\gamma-1}\frac{\partial
}{\partial r}\left[  \frac{f(\frac{r}{a(t)})}{a(t)^{N}}\right]  \\[0.1in]
&  -\kappa\theta\left[  \frac{f(\frac{r}{a(t)})}{a(t)^{N}}\right]  ^{\theta
-1}\frac{\partial}{\partial r}\left[  \frac{f(\frac{r}{a(t)})}{a(t)^{N}%
}\right]  2m(mt+n)^{-1}+\rho\frac{\delta\alpha(N)}{r^{N-1}}\int_{0}^{r}%
\frac{f(\frac{r}{a(t)})}{a(t)^{N}}s^{N-1}ds,
\end{align}
with $a(t)=(mt+n)^{2/N}.$\newline Then, we have:%
\begin{align}
&  =\rho\frac{2\left(  2-N\right)  m^{2}r}{N^{2}(mt+n)^{2}}+\gamma K\left[
\frac{f(\frac{r}{a(t)})}{a(t)^{N}}\right]  _{{}}^{\gamma-1}\frac{\dot{f}%
(\frac{r}{a(t)})}{a(t)^{N+1}}\\[0.1in]
&  -2m\kappa\theta\left[  \frac{f(\frac{r}{a(t)})}{a(t)^{N}}\right]
^{\theta-1}\frac{\dot{f}(\frac{r}{a(t)})}{a(t)^{N+2}}(mt+n)^{(2-N)/N}%
+\rho\frac{\delta\alpha(N)}{r^{N-1}}\int_{0}^{r}\frac{f(\frac{r}{a(t)}%
)}{a(t)^{N}}s^{N-1}ds\\[0.1in]
&  =\rho\left\{
\begin{array}
[c]{c}%
\frac{2\left(  2-N\right)  m^{2}r}{N^{2}(mt+n)^{2}}+\frac{K\gamma f(\frac
{r}{a(t)})^{\gamma-2}\dot{f}(\frac{r}{a(t)})}{a(t)^{N(\gamma-2)+N+1}}\\
-\frac{2m\kappa\theta f(\frac{r}{a(t)})^{\theta-2}\dot{f}(\frac{r}{a(t)}%
)}{a(t)^{N\left(  \theta-2\right)  +N+2}}(mt+n)^{(2-N)/N}+\frac{\delta
\alpha(N)}{r^{N-1}}\int_{0}^{r}\frac{f(\frac{r}{a(t)})}{a(t)^{N}}s^{N-1}ds
\end{array}
\right\}  \\[0.1in]
&  =\frac{\rho}{(mt+n)^{2(N-1)/N}}\left\{
\begin{array}
[c]{c}%
\frac{2\left(  2-N\right)  m^{2}}{N^{2}}\frac{(mt+n)^{2(N-1)/N}}{(mt+n)^{2}%
}r+\frac{\gamma Kf(\frac{r}{a(t)})^{\gamma-2}\dot{f}(\frac{r}{a(t)}%
)}{a(t)^{N(\gamma-2)+2}}\\
-\frac{2m\kappa\theta f(\frac{r}{a(t)})^{\theta-2}\dot{f}(\frac{r}{a(t)}%
)}{a(t)^{N\left(  \theta-2\right)  +3}}(mt+n)^{s-1}+\frac{\delta\alpha
(N)}{\frac{r^{N-1}}{a(t)^{N-1}}}\int_{0}^{r}\frac{f(\frac{r}{a(t)})}{a(t)^{N}%
}s^{N-1}ds
\end{array}
\right\}  \\[0.1in]
&  =\frac{\rho}{a(t)^{N-1}}\left\{
\begin{array}
[c]{c}%
\frac{2\left(  2-N\right)  m^{2}}{N^{2}}\frac{r}{(mt+n)^{N/2}}+\frac{\gamma
Kf(\frac{r}{a(t)})^{\gamma-2}\dot{f}(\frac{r}{a(t)})}{a(t)^{N(\gamma-2)+2}}\\
-\frac{2m\kappa\theta f(\frac{r}{a(t)})^{\theta-2}\dot{f}(\frac{r}{a(t)}%
)}{(mt+n)^{s(N\left(  \theta-2\right)  +3)}}(mt+n)^{s-1}+\frac{\delta
\alpha(N)}{\frac{r^{N-1}}{a(t)^{N-1}}}\int_{0}^{r/a(t)}\frac{f(s)}{a(t)^{N}%
}s^{N-1}ds
\end{array}
\right\}  \\[0.1in]
&  =\frac{\rho}{a(t)^{N-1}}\left\{
\begin{array}
[c]{c}%
\frac{2\left(  2-N\right)  m^{2}}{N^{2}}\frac{r}{a(t)}+\frac{\gamma
Kf(\frac{r}{a(t)})^{\gamma-2}\dot{f}(\frac{r}{a(t)})}{a(t)^{N(\gamma-2)+2}}\\
-2m\kappa\theta f(\frac{r}{a(t)})^{\theta-2}\dot{f}(\frac{r}{a(t)}%
)+\frac{\delta\alpha(N)}{\frac{r^{N-1}}{a(t)^{N-1}}}\int_{0}^{r/a(t)}%
\frac{f(s)}{a(t)^{N}}s^{N-1}ds
\end{array}
\right\}
\end{align}
with $\gamma=\frac{2N-2}{N}$ and $\theta=\frac{3N-4}{2N}$ for $N\geq
2.$\newline and with the new variable $z:=r/a(t)$:%
\begin{equation}
=\frac{\rho}{a(t)^{N-1}}\left\{  \frac{2(2-N)m^{2}}{N^{2}}z+\gamma
Kf(z)^{\gamma-2}\dot{f}(z)-2m\kappa\theta f(z)^{\theta-2}\dot{f}%
(z)+\frac{\delta\alpha(N)}{z^{N-1}}\int_{0}^{z}f(s)s^{N-1}ds\right\}  .
\end{equation}
In the theorem, we require the ordinary differential equation for $f(z)$:
\begin{equation}
\left\{
\begin{array}
[c]{c}%
\gamma Kf(z)^{\gamma-2}\dot{f}(z)-2m\kappa\theta f(z)^{\theta-2}\dot
{f}(z)+\frac{\delta\alpha(N)}{z^{N-1}}\int_{0}^{z}f(s)s^{N-1}ds=\frac
{2(N-2)m^{2}}{N^{2}}z,\\
y(0)=\alpha>0.
\end{array}
\right.
\end{equation}
Therefore, our solutions satisfy the momentum equation.

(2) for $\delta\Lambda>0$ and $\gamma=\theta=\frac{2N-2}{N}$, we have:%
\begin{align}
&  \rho(V_{t}+V\cdot V_{r})+K\left(  \rho^{\gamma}\right)  _{r}-\kappa
{\normalsize (}\rho^{\theta})_{r}\left(  \frac{N-1}{r}V+V_{r}\right)
-\kappa\rho_{r}(V_{rr}+\frac{N-1}{r}V_{r}-\frac{N-1}{r^{2}}V)\\[0.1in]
&  +\frac{\delta\alpha(N)\rho}{r^{N-1}}\int_{0}^{r}(\rho s^{N-1}%
-\Lambda)ds\\[0.1in]
&  =\rho\frac{\delta\alpha(N)\Lambda}{N}r+K\left(  \left[  \frac{f(\frac
{r}{a(t)})}{a(t)^{N}}\right]  ^{\gamma}\right)  _{r}-\left(  \kappa\left[
\frac{f(\frac{r}{a(t)})}{a(t)^{N}}\right]  ^{\theta}\right)  _{r}\frac
{N\dot{a}(t)}{a(t)}\\[0.1in]
&  +\rho\frac{\delta\alpha(N)}{r^{N-1}}\int_{0}^{r}\frac{f(\frac{r}{a(t)}%
)}{a(t)^{N}}s^{N-1}ds-\rho\frac{\delta\alpha(N)\Lambda}{r^{N-1}}\int_{0}%
^{r}s^{N-1}ds.
\end{align}
With $a(t)=e^{\sqrt{\frac{\delta\alpha(N)\Lambda}{N}}t},$ then we get:%
\begin{align}
&  =K\gamma\left[  \frac{f(\frac{r}{a(t)})}{a(t)^{N}}\right]  _{{}}^{\gamma
-1}\frac{\dot{f}(\frac{r}{a(t)})}{a(t)^{N+1}}-\kappa\theta\left[
\frac{f(\frac{r}{a(t)})}{a(t)^{N}}\right]  ^{\theta-1}\frac{\dot{f}(\frac
{r}{a(t)})}{a(t)^{N+1}}\sqrt{N\delta\alpha(N)\Lambda}+\rho\frac{\delta
\alpha(N)}{r^{N-1}}\int_{0}^{r}\frac{f(\frac{r}{a(t)})}{a(t)^{N}}%
s^{N-1}ds\\[0.1in]
&  =\rho\left\{  \gamma K\left[  \frac{f(\frac{r}{a(t)})}{a(t)^{N}}\right]
_{{}}^{\gamma-2}\frac{\dot{f}(\frac{r}{a(t)})}{a(t)^{N+1}}-\kappa\theta
\sqrt{N\delta\alpha(N)\Lambda}\left[  \frac{f(\frac{r}{a(t)})}{a(t)^{N}%
}\right]  ^{\theta-2}\frac{\dot{f}(\frac{r}{a(t)})}{a(t)^{N+1}}+\frac
{\delta\alpha(N)}{r^{N-1}}\int_{0}^{r}\frac{f(\frac{r}{a(t)})}{a(t)^{N}%
}s^{N-1}ds\right\}  \\[0.1in]
&  =\rho\left\{  \frac{\gamma Kf(\frac{r}{a(t)})^{\gamma-2}\dot{f}(\frac
{r}{a(t)})}{a(t)^{N(\gamma-2)+N+1}}-\frac{\kappa\theta\sqrt{N\delta
\alpha(N)\Lambda}f(\frac{r}{a(t)})^{\theta-2}\dot{f}(\frac{r}{a(t)}%
)}{a(t)^{N\left(  \theta-2\right)  +N+1}}+\frac{\delta\alpha(N)}{r^{N-1}}%
\int_{0}^{r}\frac{f(\frac{r}{a(t)})}{a(t)^{N}}s^{N-1}ds\right\}  \\[0.1in]
&  =\frac{\rho}{a(t)^{N-1}}\left\{  \frac{\gamma Kf(\frac{r}{a(t)})^{\gamma
-2}\dot{f}(\frac{r}{a(t)})}{a(t)^{N(\gamma-2)+2}}-\frac{\kappa\theta
\sqrt{N\delta\alpha(N)\Lambda}f(\frac{r}{a(t)})^{\theta-2}\dot{f}(\frac
{r}{a(t)})}{a(t)^{N\left(  \theta-2\right)  +2}}+\frac{\delta\alpha(N)}%
{\frac{r^{N-1}}{a(t)^{N-1}}}\int_{0}^{r}\frac{f(\frac{r}{a(t)})}{a(t)^{N}%
}s^{N-1}ds\right\}  \\[0.1in]
&  =\frac{\rho}{a(t)^{N-1}}\left\{  \frac{\gamma Kf(\frac{r}{a(t)})^{\gamma
-2}\dot{f}(\frac{r}{a(t)})}{a(t)^{N(\gamma-2)+2}}-\frac{\kappa\theta
\sqrt{N\delta\alpha(N)\Lambda}f(\frac{r}{a(t)})^{\theta-2}\dot{f}(\frac
{r}{a(t)})}{a(t)^{N(\gamma-2)+2}}+\frac{\delta\alpha(N)}{\frac{r^{N-1}%
}{a(t)^{N-1}}}\int_{0}^{r/a(t)}\frac{f(s)}{a(t)^{N}}s^{N-1}ds\right\}
\\[0.1in]
&  =\frac{\rho}{a(t)^{N-1}}\left\{  \gamma Kf(\frac{r}{a(t)})^{\gamma-2}%
\dot{f}(\frac{r}{a(t)})-\kappa\theta\sqrt{N\delta\alpha(N)\Lambda}f(\frac
{r}{a(t)})^{\theta-2}\dot{f}(\frac{r}{a(t)})+\frac{\delta\alpha(N)}%
{\frac{r^{N-1}}{a(t)^{N-1}}}\int_{0}^{r/a(t)}\frac{f(s)}{a(t)^{N}}%
s^{N-1}ds\right\}  ,
\end{align}
with $\gamma=\theta=\frac{2N-2}{N}$.\newline We require the corresponding
ordinary differential equations $f(z)$ with $z:=r/a(t):$%
\begin{equation}
\left\{
\begin{array}
[c]{c}%
\left(  \gamma K-\kappa\theta\sqrt{N\delta\alpha(N)\Lambda}\right)
f(z)^{\gamma-2}\dot{f}(z)+\frac{\delta\alpha(N)}{z^{N-1}}\int_{0}%
^{z}f(s)s^{N-1}ds=0\\
y(0)=\alpha>0.
\end{array}
\right.
\end{equation}
The proof is completed.
\end{proof}

\begin{remark}
If the complex number solutions $(\rho,u)\in C^{N+1}$ is considered, we have
the corresponding solutions for $\delta\Lambda<0$:%
\begin{equation}
\left\{
\begin{array}
[c]{c}%
\rho(t,r)=\frac{f(\frac{r}{a(t)})}{a(t)^{N}},\text{ }V(t,r)=i\sqrt
{\frac{-\delta\Lambda}{N}}r,\\
a(t)=e^{i\sqrt{\frac{-\delta\Lambda}{N}}t},\\
\left(  \gamma K-i\kappa\theta\sqrt{-N\delta\alpha(N)\Lambda}\right)
f(z)^{\gamma-2}\dot{f}(z)+\frac{\delta\alpha(N)}{z^{N-1}}\int_{0}%
^{z}f(s)s^{N-1}ds=0,\text{ }f(0)=\alpha,
\end{array}
\right.
\end{equation}
where $i$ is the complex constant.
\end{remark}

\begin{remark}
Our method can be easily extended to the corresponding systems with linear
damping:%
\begin{equation}
\left\{
\begin{array}
[c]{rl}%
{\normalsize \rho}_{t}{\normalsize +\nabla\cdot(\rho\vec{u})} &
{\normalsize =}{\normalsize 0,}\\
{\normalsize \rho\lbrack\vec{u}}_{t}+\left(  {\normalsize \vec{u}\cdot\nabla
}\right)  {\normalsize \vec{u}]+\beta\rho\vec{u}+\nabla P} & {\normalsize =-}%
{\normalsize \delta\rho\nabla\Phi+vis(\rho,\vec{u}),}\\
{\normalsize \Delta\Phi(t,x)} & {\normalsize =\alpha(N)}{\normalsize \rho
-\Lambda,}%
\end{array}
\right.
\end{equation}
where $\beta>0$ is a constant.
\end{remark}

\end{document}